\documentclass[12pt]{article}
\title{Comment on ``Critical assessment of the Schr\"odinger 
picture of quantum mechanics"}

\author{Hrvoje Nikoli\'c \\
Theoretical Physics Division, Rudjer Bo\v{s}kovi\'{c} Institute, \\
P.O.B. 180, HR-10002 Zagreb, Croatia \\
{\normalsize hrvoje@thphys.irb.hr} \\
\makebox[1in]{} \\
}
\date{\today}
\begin{document}
\maketitle
\begin{abstract}
Recently, Faria {\it et al} 
[Phys.~Lett.~A 305 (2002) 322]
discussed an example 
in which the Heisenberg and the Schr\"odinger pictures 
of quantum mechanics gave different results. 
We identify the mistake in their reasoning and conclude 
that the example they discussed 
does not support the inequivalence 
of these two pictures. 
\end{abstract}
\vspace*{0.5cm}
PACS: 03.65.Ta \\
Keywords: Foundations of quantum mechanics 
\vspace*{0.5cm}

\noindent

A long time ago, Dirac argued \cite{dir} that, in systems with 
an {\em infinite} number of the degrees of freedom, 
the Schr\"odinger picture of quantum mechanics 
may not be equivalent to the Heisenberg picture. The standard general 
proof of the equivalence of these two pictures may fail to be 
valid because, when an infinite number of the degrees of freedom 
is present, the formal unitary transformation between these two 
pictures may not exist in a rigorous sense. In \cite{dir}, 
Dirac argued that it was the Heisenberg picture that was 
correct in the case of an infinite number of the degrees of freedom, 
while the Schr\"odinger picture was wrong. The two pictures 
are really equivalent only when the number of the degrees of 
freedom is finite.
 
Recently, Faria {\it et al} \cite{far} considered an example 
in which they explicitly obtained a discrepancy between the 
results obtained in the two pictures. Specifically, 
they considered a charged harmonic oscillator in 3 dimensions 
(having 3, i.e. a {\em finite} number of the degrees of freedom!) 
interacting with the electromagnetic field (having an infinite 
number of the degrees of freedom). For simplicity, they 
studied only the $x$-direction of the harmonic oscillator.
In the Heisenberg picture, 
they quantized both the harmonic oscillator and the 
electromagnetic field and obtained that the average square of the 
position operator of the harmonic oscillator in the ground state 
is equal to 
$\langle x^2\rangle =\hbar/2m\omega_0$ (where $m$ is the mass 
of the harmonic oscillator and $\omega_0$ is its frequency), 
the same result as if the interaction with the electromagnetic field 
was absent. In the Schr\"odinger picture, they used a semiclassical
approximation, i.e. they quantized only 
the harmonic oscillator, while  
they treated the electromagnetic field as a classical field. 
Using such a Schr\"odinger picture, 
they obtained $\langle x^2\rangle =\hbar/m\omega_0$, i.e.
twice the value of that in the Heisenberg picture. 
From this result they concluded that
the result obtained in the Heisenberg picture was correct, while that
obtained in the Schr\"odinger picture was wrong. Consequently, 
they concluded that the two pictures were not equivalent.     
 
However, two objections to their conclusions seem natural.
First, they used the semiclassical approximation only in one 
of the pictures, so in general one does not expect the results 
obtained in the two pictures to be identical. Therefore, the 
discrepancy they obtained may have nothing to do with 
a possible inequivalence of the two pictures.  
Second, in the Schr\"odinger picture they quantized only a 
finite number of the degrees of freedom, in which case 
the Schr\"odinger picture should be correct. 
Therefore, their results cannot be taken as an indication 
that the two pictures are inequivalent.
In order to see more closely where they made a mistake in their 
reasoning, we make their reasoning more transparent 
by reviewing the main steps in their analysis and using a 
slightly different notation. 

In the Heisenberg picture, the ground-state expected value of square 
of the position operator is 
\begin{equation}\label{e1}
\langle x^2 \rangle_H (t)= \langle \phi_0 | \hat{x}_H^2(t) | 
\phi_0 \rangle,
\end{equation}
where $|\phi_0\rangle$ represents the ground state.
The explicit value of the right-hand side of (\ref{e1}) is given by 
(14') (where the prime denotes equations in \cite{far}).
 
In the Schr\"odinger picture, they show that the ground-state wave
function takes the form
\begin{equation}\label{e2}
\psi(x,t)=\phi_0 (x-q_c(t)) e^{i\varphi(t)},
\end{equation} 
where $\phi_0 (x)$ is the normalized ground-state wave function of a 
``free" (i.e. without the electromagnetic interaction) 
harmonic oscillator and $q_c(t)$ is a c-number function that 
satisfies the classical equation of motion
\begin{equation}\label{e3}
m\ddot{q}_c(t)=-m\omega_0^2 q_c(t) +eE(t).
\end{equation}
Here $E(t)$ is the classical electric field in the $x$-direction.
It is the sum of the classical vacuum field and the classical 
radiation reaction field.  
The expected value of square of the position operator is 
\begin{equation}\label{e4}
\langle x^2 \rangle_S (t)= \int_{-\infty}^{\infty}
dx\, \psi^*(x,t)x^2\psi(x,t).
\end{equation}
After making a shift of the integration variable 
$x\rightarrow x-q_c(t)$ and using (\ref{e2}), one finds
\begin{equation}\label{e5}                                  
\langle x^2 \rangle_S (t)= \int_{-\infty}^{\infty}    
dx\, \phi_0^*(x)x^2\phi_0(x) +
\int_{-\infty}^{\infty}dx\, \phi_0^*(x)q^2_c(t)\phi_0(x). 
\end{equation}
This result corresponds to (30').

So far so good. However, now comes their step that turns out 
to be wrong. {\it From the fact that $q_c(t)$ and 
$\langle \phi_0 | \hat{x}_H(t) |\phi_0 \rangle$ obey the same 
equation of motion (\ref{e3}), they conclude that the second term 
in (\ref{e5}) is equal to the right-hand side of (\ref{e1}), 
i.e. that} 
\begin{equation}\label{e5.1}
\langle \phi_0 | \hat{x}_H^2(t) |\phi_0 \rangle=
q_c^2(t).
\end{equation}
It is clear that such a reasoning is not correct, i.e. that, 
in general, 
(\ref{e5.1}) does not have to be valid. Nevertheless, with such a
reasoning, (\ref{e5}) becomes  
\begin{equation}\label{e6}    
\langle x^2 \rangle_S (t)= \int_{-\infty}^{\infty}
dx\, \phi_0^*(x)x^2\phi_0(x) +
\langle x^2 \rangle_H (t),
\end{equation}
which corresponds to (31'). From this, they concluded that
the Schr\"odinger picture was not equivalent
to the Heisenberg picture, i.e. that the Schr\"odinger picture 
led to the wrong result (\ref{e6}).

We agree that (\ref{e6}) is wrong, because if the 
two pictures are equivalent, then 
$\langle x^2 \rangle_S (t)=\langle x^2 \rangle_H (t)$. 
However, we do not agree that the wrong result
(\ref{e6}) implies that  
the Schr\"odinger picture is wrong. Instead, it is their 
reasoning designated by italics above that is wrong.

Note that if the reasoning designated by italics was correct, 
then one would obtain (\ref{e6})  
for any value of $e$ in (\ref{e3}), including the case 
$e\rightarrow 0$. This would imply that the 
two pictures are not equivalent even without the electromagnetic 
interaction. On the other hand, it is well known 
that the two pictures are equivalent for such a simple 
harmonic oscillator. Indeed, when $e\rightarrow 0$, 
one of the solutions of (\ref{e3}) is $q_c(t)=0$ for all $t$.
This solution does {\em not} satisfy (\ref{e5.1}). 
With this solution, the second term in (\ref{e5}) vanishes, 
leading to 
\begin{equation}\label{e7}
\langle x^2 \rangle_S (t)= \int_{-\infty}^{\infty}
dx\, \phi_0^*(x)x^2\phi_0(x).
\end{equation}
Contrary to (\ref{e6}), this is a consistent result.
It is shown in \cite{far} that the right-hand sides 
of (\ref{e7}) and (\ref{e1}) are {\em both} equal to
$\hbar/2m\omega_0$.

When $e\neq 0$, then the solution $q_c(t)$ of (\ref{e3}) 
does not vanish, provided that $E(t)$ does not vanish.
Consequently, the second term in (\ref{e5}) does not vanish, 
which contradicts the result obtained in the Heisenberg picture in
\cite{far}. However, in \cite{far}, the function 
$E(t)$ has not been calculated explicitly. 
A correct treatment of the case $e\neq 0$ is given in 
\cite{boy,fra}.
Since $E(t)$ in \cite{far} does not contain an external field, 
(\ref{e3}) describes a damped harmonic oscillator \cite{fra}, 
so that $q_c(t)$ eventually vanishes. This is also in agreement 
with the results of \cite{boy}. The result $q_c(t)=0$ can 
also be understood from an approach in which the electric 
field is quantized, because it appears that the two contributions 
to the electric field (the vacuum field and the radiation reaction field) 
cancel in the ground state \cite{mil}.

By reconsidering the example considered in \cite{far}, 
we have found that this example does not support 
the inequivalence of the Schr\"odinger picture 
with the Heisenberg picture. In this way, we have shown 
that the main conclusion of \cite{far} is incorrect.
Of course, the general arguments against the Schr\"odinger picture
\cite{dir} mentioned in the introductory paragraph are 
not discredited by our result. However, we note that 
today the formal problem with the Schr\"odinger picture 
of quantum field theory is usually not 
considered as a serious problem because today theorists 
are used to use various methods of regularization 
that cure the pathologies emerging from theories with 
an infinite number of the degrees of freedom.
For example, the regularization on a lattice transforms the
theory with an infinite number of the degrees of freedom
into a theory with a finite number of them, in which case 
the two pictures are equivalent.

The author is grateful to an anonymous referee
for the valuable remarks.
This work was supported by the Ministry of Science and Technology of the
Republic of Croatia under Contract No. 0098002.


\begin{thebibliography}{3}

\bibitem{dir}
P.A.M.~Dirac, Phys.~Rev.~139 (1965) B684.
\bibitem{far}
A.J.~Faria, H.M.~Franca, C.P.~Malta, R.C.~Sponchiado, 
Phys.~Lett.~A 305 (2002) 322.
\bibitem{boy}
T.H.~Boyer, Phys.~Rev.~D 11 (1975) 790.
\bibitem{fra}
H.M.~Franca, M.T.~Thomaz, Phys.~Rev.~D 31 (1985) 1337;
Erratum 38 (1988) 2651.
\bibitem{mil}
P.W.~Milonni, Am.~J.~Phys.~52 (1984) 340.

\end{thebibliography}
\end{document}